\documentclass[twocolumn,showpacs,preprintnumbers,amsmath,amssymb]{revtex4}

\usepackage{graphicx}
\usepackage{dcolumn}
\usepackage{bm}

\begin{document}

\preprint{APS/123-QED}

\title{Construction and Analysis of Random Networks with Explosive Percolation}

\author{Eric J. Friedman$^1$ and Adam S. Landsberg$^2$\\
\em
$^1$School of ORIE and Center for Applied Mathematics, Cornell University, Ithaca, NY 14850.\\
$^2$Joint Science Department, Claremont McKenna, Pitzer, and Scripps Colleges, Claremont, CA 91711}%

\date{\today}

\begin{abstract}
The existence of explosive  phase transitions in random (Erd\H os  R\'enyi-type) networks has been recently documented by Achlioptas et al.\  [Science {\bf 323}, 1453 (2009)] via simulations. In this Letter we describe the underlying mechanism behind these first-order phase transitions and develop tools that allow us to identify (and predict) when a random network will exhibit an explosive transition. Several interesting new  models displaying explosive transitions are also presented.
\end{abstract}

\pacs{Valid PACS appear here}
\maketitle

The structure and dynamics of networked models and their application to social networks is an important and active area of research encompassing many fields ranging from physics \cite{AlB02,Dor02, Coh03} to sociology \cite{Wat03} to combinations thereof \cite{NSW01,NBW06,WaS98}. Ideas from statistical mechanics have contributed greatly to our understanding of such networks and their practical uses \cite{CEA01,Bar03, New04}. Of particular importance is the statistical mechanical notion of the order of a phase (`percolation') transition. Phase transitions in random network models are almost always second order \cite{ErR60,CHK01} or higher \cite{CHK01, KKK02}. Thus, it was surprising to many when Achlioptas et al.\  \cite{ADS09} reported recently that some models of interest in social networks can display first-order (discontinuous) transitions.

In that work, they described several random graph models of the Erd\H os  R\'enyi (ER) variety that exhibit first-order or what they call ``explosive'' phase transitions.  They provide convincing numerical evidence and a useful characterization of such transitions, but no details on the mechanisms underlying them. They describe several systems which display such transitions and a general class of systems which don't.

In this paper we describe the underlying mechanisms behind explosive transitions in ER-type models.  We show that, somewhat surprisingly, the key to explosive transitions  is not the details of the edge-addition rules at work during the actual ``explosion,'' but rather lies in the period preceding the explosion when a type  of ``powder keg''  develops.   In effect, the importance of the rules is to create an explosive situation, which can be detonated with almost any rule.
In addition, our analysis provides an understanding of which random network models will have such transitions. This allows us to construct large classes of interesting models that display this behavior. (It also allows us to rule out many other models which will not display explosive transitions.)

The prototypical network percolation example is that of pure (non-preferential) ER random graphs \cite{ErR60}. These begin with a set of $n$ nodes, where $n$ is large. Edges are then added to the graph, uniformly at random.
As is well known, this system exhibits a phase transition as the number of edges $\tau$ increases. For $\tau<0.5n$ all clusters are small ($\sim log(n)$) while for $\tau>0.5n$ a large cluster ($\sim n$) appears.  In the large-$n$ limit this transition is a second-order phase transition, i.e., letting $s(\tau)$ be the size of the largest cluster after $\tau$ edges have been added, the graph of $s(\tau)/n$ against $\tau/n$ is continuous, as seen in Figure~1.

\begin{figure}
\includegraphics[scale=.25]{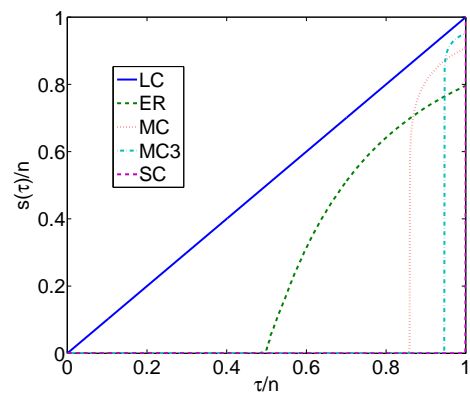}
\caption{Transition diagrams for largest cluster (LC), Erdos-Renyi (ER), min-cluster (MC), min-cluster 3 (MC3) and smallest cluster (SC) rules.
LC and ER display second-order transitions while the others are first order.}
\end{figure}

Achlioptas et al.\   \cite{ADS09} considered a variety of ER-like random networks using modified edge-addition procedures, wherein, at every step, two candidate edges are chosen at random, but only one of the two is actually added to the graph. Under the min-cluster (MC) rule, for example, one  selects the  edge  which minimizes the sum of the weights of the nodes in the edge, thereby creating the smaller cluster. (Here, the ``weight'' of a node is defined as the size of the connected component (cluster) which contains that node.)  As mentioned in \cite{ADS09},  this rule  leads to explosive phase transitions, as evidenced by the discontinuity in a plot of $s(\tau)/n$ against $\tau/n$ (Figure~1). To demonstrate that such transitions are truly first order (discontinuous), they use the following approach:  Letting $t(a)$ be the `time' (i.e., number of edges added) when a cluster of size $ \geq a$ first arises, define $\Delta = t(n/2)-t(n^{\alpha})$, for some $0<\alpha<1$.  Then, to show that a transition is explosive they demonstrate,  via simulations, that $\Delta \sim n^{\beta}$ for some $\beta<1$  and thus the width of the transition region in the rescaled network is $\Delta/n \sim n^{\beta-1}$, which vanishes in the large $n$ limit. For the MC model, when $\alpha=1/2$ we find that $\beta\approx 0.6$, as shown in Figure~2. (Note that the condition $\beta<1$ only implies first order transitions in ER-type models, which include Achlioptas processes. This condition is not sufficient for some other classes of models \cite{Zif09,CKP09,RaF09}.)

\begin{figure}
\includegraphics[scale=0.28]{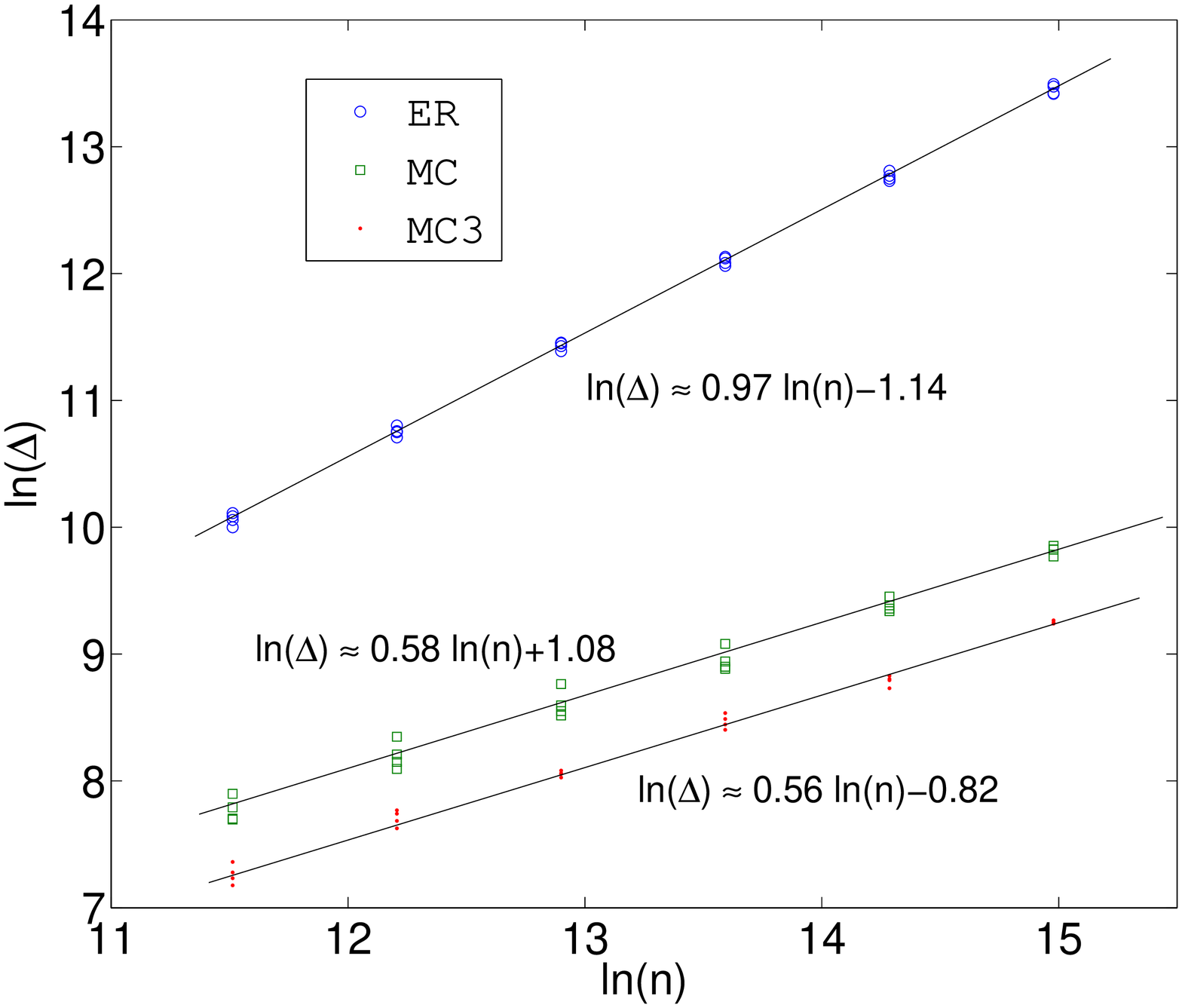}
\caption{A log-log plot of $\Delta$  vs. $n$ for ER, MC and MC3 rules, for $\alpha=1/2$. (Markers are for individual realizations.) }
\end{figure}

Other edge addition rules were also considered by Achlioptas et al.\   For instance, under the product rule (PR), one selects whichever of the two candidate edges has the smaller product of its node weights.  This too leads to an explosive transition.  On the other hand, adding the edge whose sum (or product) of its node weights is largest does not.  Their numerical work suggests
that edge-addition rules which favor the formation of smaller clusters are somehow linked to explosive transitions, though no formal analysis has been provided nor have the underlying mechanisms for generating explosive transitions been identified.

Our first contribution to understanding explosive transitions in random networks is the insight that the key to the explosion is the special preparation that occurs beforehand. Let $F(\tau,a)$ be the number of nodes in clusters of size $\geq a$ after the addition of the $\tau$'th edge.  We find that, under the MC  rule, $$\frac{1}{n}F({t(n^{\alpha})},n^{1-\beta})$$ approaches a non-zero constant in the large-$n$ limit, as shown in Figure~3 for $\alpha=1/2$ and $\beta = 0.6$.
In other words, at time $t(n^{\alpha})$, which is the beginning of the phase transition, a fixed (non-zero) fraction of nodes are contained in small clusters with sizes ranging from $n^{1-\beta}$ to $n^{\alpha}$.
The set of clusters in this size range constitutes what we term the ``powder keg.''  Although each individual cluster in the powder keg contains only a vanishingly small fraction of nodes, they are ignitable and {\em collectively} they enable an explosion to occur.
Note that an analogous plot for the ER rule (Figure~3) shows that in that case the powder keg is empty in the large $n$ limit.
\begin{figure}
\includegraphics[scale=0.23]{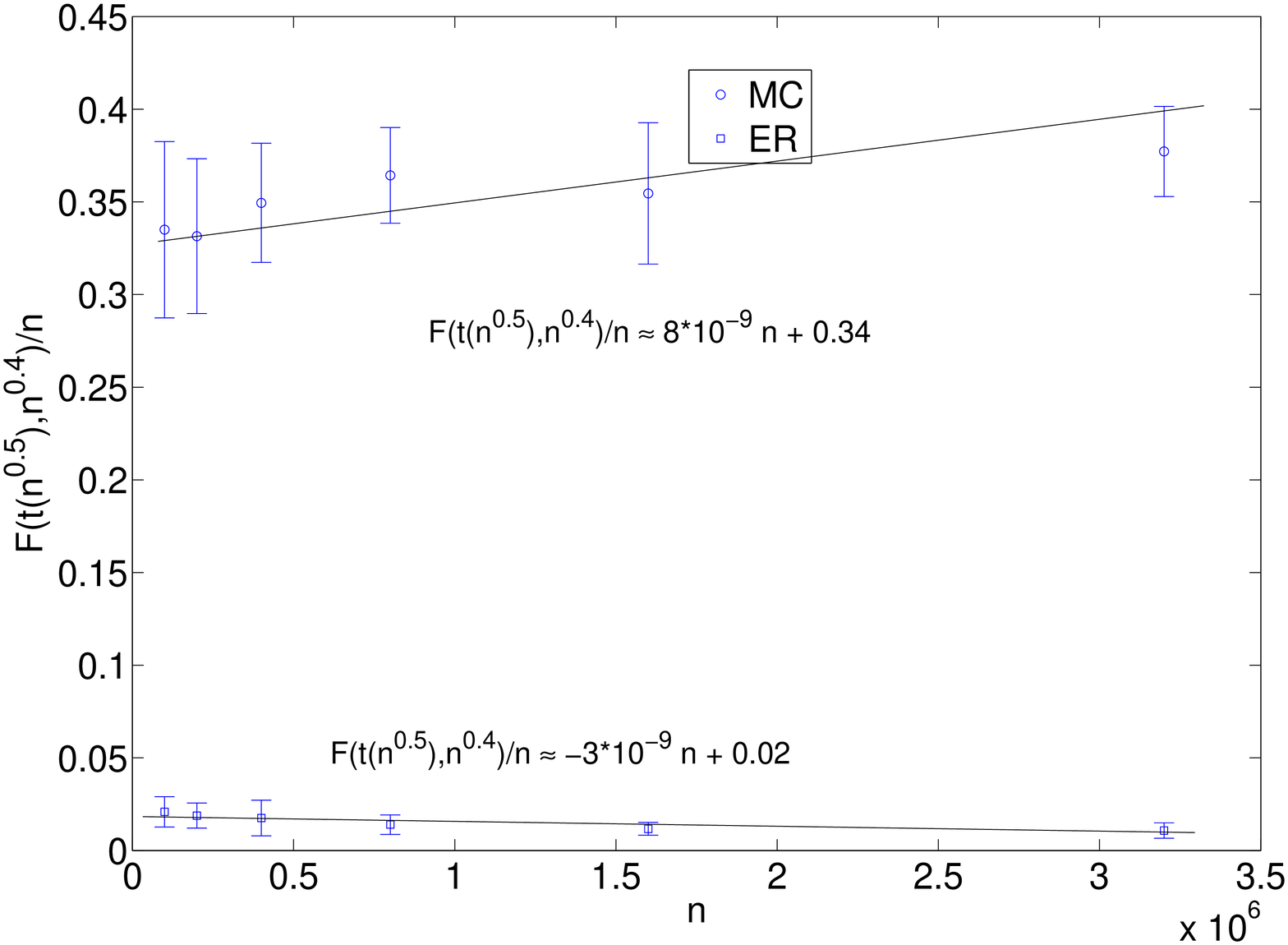}
\caption{Plot of $\frac{1}{n}F({t(n^{0.5})},n^{0.4})$ as a function of $n$ for the Erdos-Renyi (ER) and min-cluster (MC) rules. All coefficients of regression are statistically significant ($p<10^{-3}$).}
\end{figure}

To see why the existence of a powder keg guarantees an eventual explosion we first make a simple observation: If there is no cluster of size $\geq n/2$ in the network, then the probability of an edge being added which connects two nodes within the same cluster is less than some fixed constant strictly less than 1. For example, in the ER rule this constant is $1/2$ while for the MC rule this constant is $1/4$. Consequently, we see that internal edges in the graph will not impact the order of the phase transition itself, and hence can be safely ignored. In light of this, we define  $\hat{\tau}$ to be a measure of the number of ``non-internal'' edges added to the graph.  In general  if there are $C$ clusters in total in a graph at non-internal time $\hat{\tau}$, then at (non-internal) time $\hat{\tau} + C-1$ there will be a single giant cluster containing all nodes, since each new edge must join two clusters together.

Now, for the case of the MC rule, focus on the clusters of size $\geq n^{1-\beta}$ which exist at time $t(n^{\alpha})$, i.e., clusters in the powder keg.  Note that there can be at most $n^{\beta}$ such clusters, and that a finite (non-zero) fraction of nodes in the network belong to the powder keg,  as shown in Figure~3.  Thus, the probability of choosing an edge that connects two different clusters, {\em both in the powder keg}, is strictly greater than 0. It therefore follows that the time from the creation of the powder keg, $t(n^{\alpha})$, to the time when a cluster of size $n/2$ first forms, $t(n/2)$, is at most $\sim n^{\beta}$, since a positive fixed  fraction of the edges added joined clusters from the powder keg. Thus $$\Delta =  t(n/2)-t(n^{\alpha}) \sim n^{\beta}.$$    So once a powder keg is created, any reasonable edge-addition rule will eventually detonate it -- i.e., {\em the existence of a powder keg guarantees an eventual explosive transition in the network}.

To better understand this difference between non-explosive models (e.g., pure ER) and explosive models (e.g., MC, PR), it is helpful to make note of two very extreme, highly simplified models for which  analytical calculations are possible.  Surprisingly, despite their simplicity, these capture many of the essential distinctions between second- and first-order transitions in networks. In the so-called largest cluster (LC) model,  at each step one identifies the two largest clusters in the network and adds an edge between them. In this case the dynamics is characterized by a single cluster growing over time, and a straightforward computation shows that $\Delta \approx (n/2-1) - (n^{1/2}-1)$. Since $\Delta$ is linear in $n$ the transition will be non-explosive. (See Figure~1.)  In the smallest cluster (SC) model, at each step one identifies the two smallest clusters in the network and adds an edge between them. In this case a  calculation shows that $t(a) \approx n(1-2/a),$ since the first $n/2$ steps simply create $n/2$ clusters of size 2 while the next $n/4$ steps create $n/4$ clusters of size $4$ and so on. Thus  $\Delta \approx (n-4)-(n-2n^{1/2}).$ Hence $\Delta \sim n^{1/2}$, which is explosive. In fact the SC rule is essentially the ``most explosive'' rule and more explosive than the MC or PR rules. Its (rescaled) graph is simply a step function at 1. (See Figure~1.) Note that in this model the powder keg is extreme, as all the nodes are always in the interval  $[n^{\alpha}/2,n^{\alpha}]$ at $t(n^{\alpha})$, if $n^\alpha$ is a power of 2. This implies that $\alpha+\beta = 1$, whereas $\alpha+\beta >1$ for the MC model.

Having established earlier that a powder keg guarantees an eventual explosion in the network, we now turn to the question of how a powder keg comes into being.  We first focus on the process by which a new edge creates a cluster of size $\geq a$.  Recall that under the MC rule, one considers two potential edges (randomly chosen) and selects the one leading to the smaller resultant cluster size. So
in order to form a cluster of size  $\geq a$, both these potential edges must contain at least one node of size $\geq a/2$ (otherwise the edge selected by the MC rule would create a cluster of size less than $a$). Hence the probability that the addition of a new edge to a graph at time $\tau$ will produce a cluster of size $\geq a$  is at most $4(F(\tau,a/2)/n)^2$ to lowest order, since $F(\tau,a/2)/n$ is the probability of randomly choosing a node of size $\geq a/2$. (Formally, the function $F$ is a random variable defined on sample paths. In our analysis, the bounds on $F$ are all  high probability bounds,  i.e., the probability of them not holding will vanish as $n\rightarrow\infty$.)

We next establish a lower bound for the quantity $F({t(a)},a/2)$, which represents the number of nodes in clusters of size  $\geq a/2$  at the time $t(a)$ when the first cluster of size $\geq a$ appears.
First, observe that the expected number of clusters of size $\geq a$ at this time (up to constants and higher-order terms) is at most
$$\sum_{i=0}^{t(a)} 4(F(i,a/2)/n)^2$$ which is bounded  by $4 t(a) (F({t(a)},a/2)/n)^2$ because $F(u,a/2)$ is non-decreasing in $u$. Next, we can write this bound as $4n (F({t(a)},a/2)/n)^2$  since the previous argument concerning non-internal time shows that only values of time which are less than $n$  need be considered.
Now suppose (incorrectly) that $F({t(a)},a/2)$ were less than $n^{1/2-\epsilon}$ for some $\epsilon>0$. Then the expected number of  clusters of size $\geq a$ at time $t(a)$ would  be vanishingly small in the large-$n$ limit ($\sim n^{-2 \epsilon}$). This contradicts the meaning of $t(a)$ as the time marking the first appearance of a cluster of size $\geq a$ in the network.
Hence, at the time of the creation of the first cluster of size $a$, it must be the case that $F({t(a)},a/2)$ is of order $n^{1/2}$ or greater.

This bound is essential for understanding the creation of the powder keg under the MC and similar rules.  In essence, as a lower bound it underpins the build-up of a sufficient number of clusters in the size range that constitutes the powder keg.  We note that the analogous computation for the ER rule yields no corresponding lower bound on $F({t(a)},a/2)$, and hence no powder keg develops.  As we will discuss later, this allows us to predict which models will (and won't) have explosive transitions.

Continuing, we can extract additional information from this lower bound by iterating the previous argument.  (For notational convenience we drop the $\epsilon$'s from the analysis, since they can be chosen to be arbitrarily small.)
Consider the situation when  a cluster of size $n^\alpha$ first forms, for $\alpha \leq 1/2$.
We just showed that, at this time, the number of nodes $F({t(n^\alpha)},n^\alpha/2)$ is of order $n^{1/2}$ or greater, which implies that there are at least order $n^{1/2-\alpha}$ clusters of size $\sim n^\alpha$. Now, the  expected number of such clusters is bounded by
$$n^{1/2-\alpha} \approx \sum_{i=0}^{t(n^\alpha)} 4(F(i,a/4)/n)^2 \leq 4n (F({t(n^\alpha)},a/4)/n)^2 $$ which implies that
$F({t(n^\alpha)},n^{\alpha}/4)$ is order $ n^{3/4-\alpha/2}.$
Iterating this  argument shows that
$F({t(n^\alpha)},n^\alpha/2^k)$ is order $n^{1-\alpha +(2 \alpha -1)/2^k}.$ This result hints at the creation of the powder keg, but is not sufficient to prove its existence.

Improving this argument requires a more detailed analysis of the dynamics of the system. The following argument is heuristic and instructive. (It remains an open problem to formalize the argument.)
Observe that no clusters of size $n^{\alpha}$ can appear until at least one cluster of size $n^{\alpha}/2$ arises. This implies that the bound on the expected number of clusters of size $n^{\alpha}$ can be refined to
$$\sum_{i=0}^{t(n^{\alpha})} (F(i,n^{\alpha}/2)/n)^2 =  \sum_{i=t(n^{\alpha}/2)}^{t(n^{\alpha})} (F(i,n^{\alpha}/2)/n)^2$$ which is bounded by\\ $[t(n^{\alpha})- t(n^{\alpha}/2)] (F(t(n^\alpha),n^{\alpha}/2)/n)^2$. Unfortunately, the analytic computation of $[t(n^{\alpha})- t(n^{\alpha}/2)]$ appears to be complex and is extremely model dependent. However,  for the SC example, this time interval is of order $n^{\beta}$, where $\beta=1-\alpha$, so the expected number of clusters is
$n^{1-\alpha}(F(t(n^\alpha),n^{\alpha}/2)/n)^2$
which yields $F(t(n^\alpha),n^{\alpha}/2)\sim n^{\frac{1+\alpha}{2}}$. Iterating this argument gives
the bound
$F({t(n^{\alpha})},n^{\alpha}/2^k) \sim n^{1-\frac{1-\alpha}{2^k}}$.  For $k$ suitably large this
indicates that the powder keg  can contain enough mass.

Our analysis provides the foundations on which to understand other ER-like models with explosive transitions. The key insight is the equation for the probability of creating a large cluster from smaller ones. In particular, we conjecture that any ``reasonable'' network model for which the probability of creating a cluster of size $a$ at time $t$ has probability proportional to $(F(t,a/2)/n)^p$ with $p>1$ will be explosive, since then extensions of the analysis above will apply.  This conjecture  provides a simple criterion for predicting if a random network will be explosive and for creating new variant networks with this property.

For example, one can generate an interesting class of models by considering $k$ randomly chosen candidate edges and selecting the one with the lightest node sum. When $k=1$ this is the ER model, which is non-explosive since $p=1$.  For $k=2$ this is the MC model which is explosive ($p>1$). For larger $k$ these models are increasingly explosive and are denoted min-cluster-$k$ rules. For example, in Figure~1 the min-cluster-3 rule (MC3)  appears ``more explosive'' than the MC rule (though Figure~2's plot of $\Delta\sim n^\beta$ is not numerically accurate enough to firmly establish that MC3 truly has a smaller $\beta$ value than that of MC).    We can also extend this class of models to choosing the m'th lightest out of $k$  for $m\leq k$ (henceforth called the $(m,k)$ models). Our analysis suggests that for $m=1$ these models are not explosive, but for $m>1$ they are (since $p>1$), as we have confirmed numerically for modest values of $k$.

One of the limitations of the explosive network models considered by Achlioptas et al.\ is that  they require a comparison between two unrelated edges which may lie in completely distinct regions of the network.   Unfortunately, since in practice random network models are often used to describe decentralized processes (e.g., growth of social or financial networks, spread of disease, etc.), such a comparison between unrelated edges can be a bit artificial in these circumstances. However, using our criterion above we can readily construct variant models in which the edges being compared have the feature that they share common nodes (which is often the case in social networks).
As an example, consider a model in which one simply picks three nodes at random in the network and chooses the lightest edge between them. A calculation shows that $p=2$ for this model and hence it is expected to be explosive (as we have numerically verified).  Another interesting model which can arise in social networks is the following:  An edge between two randomly selected nodes is proposed. This edge will only be ``accepted'' if both nodes agree to the proposed union. The nodes make their decision as follows:  Each node in the edge picks a second node at random and compares the weight of that second node with that of its original proposed partner. If the original proposed  partner node has a smaller weight than the second node then the node accepts the edge. If both nodes in the proposed edge accept it then the edge is added, otherwise it is not and the process is repeated.  A straightforward computation shows that this model has $p=2$ and thus is expected to exhibit an explosive phase transition (which we have verified numerically). Many other such variations are also possible, and can be readily classified using our analysis.  For example, if one were to modify the previous model by having only one of the nodes in the proposed edge do a comparison, then a quick check would reveal that $p=0$, leading one to predict that the transition will not be explosive (as has been numerically verified).  Likewise, one can show that most models in which the heaviest edge is chosen will not exhibit explosive transitions.

As a final observation, we note that in this paper we have identified the underlying mechanism responsible for explosive transitions in the ER-type networks studied by Achlioptas et al.\, and provided a criterion that appears to offer direct guidance for easily recognizing and predicting whether or not a given random network will display an explosive transition.  This in turn may prove helpful in designing new random ER-type network models with desired characteristics, as in some situations explosive transitions are desired and in others not.  Interestingly, recent observations of explosive transitions have been reported by Ziff \cite{Zif09} for  two-dimensional percolation models, and by Cho et al.\ \cite{CKP09} and Radicchi and Fortunato \cite{RaF09} for scale-free networks. It is a significant but currently open question as to how the underlying mechanism and methodology for understanding explosive transitions in ER-type models is related to the analogous transitions recently seen in these other systems.

The authors thank Seth Marvel, Joel Nishimura and Steve Strogatz for helpful conversations. EJF's research has been supported
in part by the NSF under grants ITR-0325453 and CDI-0835706.

\bibliography{../bib}

\end{document}